\documentstyle[preprint,aps,prb]{revtex}
\begin{document}

\title{NiO Exchange Bias Layers Grown by Direct Ion Beam Sputtering 
of a Nickel Oxide  Target}

\author{Richard P. Michel\cite{Remail} and Alison Chaiken\cite{Aemail}}
\address{Mailstop
L-350, Lawrence Livermore National Laboratory, Livermore, CA 94550}

\author{Young K. Kim\cite{Yemail}}
\address{Quantum Peripherals Colorado Inc.  Louisville, CO 80028-8188}

\author{Lantz E. Johnson\cite{Lemail}}
\address{Lawrence Livermore National Laboratory, Livermore, CA 94550}

\maketitle

\begin{abstract}
A new process for fabricating NiO exchange bias layers has been
developed.  The process involves the direct ion beam sputtering (IBS)
of a NiO target.  The process is simpler than other deposition
techniques for producing NiO buffer layers, and facilitates the
deposition of an entire spin-valve layered structure using IBS without
breaking vacuum.  The layer thickness and temperature dependence of
the exchange field for NiO/NiFe films produced using IBS are presented
and are similar to those reported for similar films deposited using
reactive magnetron sputtering.  The magnetic properties of highly
textured exchange couples deposited on single crystal substrates are
compared to those of simultaneously deposited polycrystalline films,
and both show comparable exchange fields.  These results are compared
to current theories describing the exchange coupling at the NiO/NiFe
interface.
\end{abstract}

\section{INTRODUCTION}

	 Antiferromagnetic (AF) NiO buffer layers used to exchange
bias ferromagnetic layers and for domain stabilization in
magnetoresistive sensors are primarily fabricated using a reactive
sputtering technique developed by Carey {\em et al.}\cite{matt,matt2}
Laser ablation,\cite{kennedy} reactive MBE\cite{lind} and
MOCVD\cite{lai} have also been employed for this purpose with some
success.  We have developed a new simpler process for depositing NiO
films using ion beam sputtering. The technique involves the direct
sputtering of a NiO target and so does not require an oxygen partial
pressure.

	The ion beam deposition chamber used to deposit the magnetic
films has been described elsewhere.\cite{fesiprb} A NiO target is
sputtered by a neutralized Ar ion beam and the ejected material
accumulates on substrates suspended 25 cm above the target.  The
deposition rate is 0.1 \AA/sec. The substrate temperature is
approximately 60$^{\circ}$C during deposition unless otherwise
noted. An uniaxial anisotropy is established in the NiFe layers with
permanent magnets which produce a 300 Oe uniform bias field, $\rm H_b$, at
the substrates during deposition.  We use high-angle x-ray
diffraction (XRD) to study the morphology of the NiO layers and low
angle XRD to calibrate the film thickness. A vibrating sample
magnetometer equipped with a high temperature oven is used to
determine the magnetic properties of the NiO/NiFe coupled
films. Rutherford backscattering measurements show that the NiO is
stoichiometric to within 1\%.

\section{EXPERIMENTAL RESULTS}

	Fig. 1a shows the easy and hard axis magnetization of a
NiO(500\AA)/NiFe(100\AA) bilayer deposited on Si with an amorphous
$\rm Al_2O_3$ film. The NiO film is untextured with a grain size in
the growth direction, derived from the FWHM of the NiO(200) Bragg
peak, of 80\AA. The easy axis loop is offset by $\rm H_E$ = 66 Oe
which corresponds to an interface exchange energy of J = 0.046
erg/cm$^2$. The largest value we observed was J = 0.065
erg/cm$^2$. The energies are in agreement with those observed by Carey
and coworkers.\cite{matt} NiCoO/NiFe, and NiO/Co bilayers with
promising exchange fields have also been fabricated using this new IBS
sputtering technique. Fig. 1b shows the magnetization of a bilayer
deposited simultaneously with the films shown in Fig. 1a, but on a
polished single crystal (001) oriented MgO substrate.  XRD shows both
the NiO and the NiFe are highly (001) textured.  The magnetization of
the NiFe shows evidence of both a induced uniaxial anisotropy along
the $\rm H_b$ axis, and unidirectional anisotropy defined by the
direction of $\rm H_b$.  Fig.  1b shows M(H) in the in-plane (100)
directions, one parallel to $\rm H_b$ and one perpendicular to $\rm
H_b$. The exchange field H$\rm _E$ is 20 Oe (J = 0.015 erg/cm$^2$).

	The dependence of H$\rm _E$ on the thickness of the IBS grown
NiO and NiFe layers was measured. Consistent with other
studies,\cite{lin} the exchange field agrees with $\rm t_{NiFe}^{-1}$
behavior for 300\AA $>$ $\rm t_{NiFe}$ $>$ 50\AA and t$\rm _{NiO}$ =
500\AA, as expected from the interfacial origin of the
interaction. The H$\rm _E$ is approximately constant for t$\rm _{NiO}$
$>$ 400\AA, and decreases to zero at a critical thickness of about
175\AA. The easy axis coercivity peaks near the critical
thickness. The optimum NiO thickness where the difference between
H$\rm _E$ and H$\rm _c$ is maximum is between 400\AA and
500\AA. Finally we find the training effect, which is the reduction of
H$\rm _E$ after multiple field cycles, is largest for NiO thicknesses
near the critical thickness. No training effect was observed for NiO
films greater than 400\AA.

The morphology of the NiO deposited on an amorphous buffer layer is
sensitive to the detailed deposition conditions. We have studied the
effect of substrate deposition temperature, beam voltage, deposition
rate and substrate voltage bias on the texture of 500\AA\ thick NiO
layers. We observe NiO (111), (200) and (220) Bragg reflections with
various intensities (the positions indicate the NiO lattice is
expanded relative to bulk by 1\%). We find, however, that H$\rm _E$ is
not sensitive to the changes in the bulk NiO morphology detected at
this level.  Though we find variations in the strength of H$\rm _E$,
they are not correlated to variation on the NiO texture.  This
conclusion is illustrated in Fig. 2 where the interface exchange
energy, J = H$\rm _E M_s t_{NiFe}$, for a wide range of
NiO(500\AA)/NiFe(t$\rm _{NiFe}$) films is plotted as a function of the
ratio of the NiO (111) to the NiO (200) Bragg peak intensities. The
data are space filling indicating that the bulk texture of the NiO
layer is not a good predictor of the strength of the interface
coupling.

  The Neel transition temperature, $\rm T_N$, for bulk NiO is
250$^{\circ}$C. The H$\rm _E$ of NiO/NiFe bilayers typically drops to
zero at a lower blocking temperature, $\rm T_b$, which also varies
with the interface and bulk morphological properties of the bilayer
couple.\cite{lin} The temperature dependence of H$\rm _E$ and H$\rm
_c$ of an IBS grown NiO 500\AA/NiFe 50\AA coupled film is shown in
Fig. 3.  H$\rm _E$ drops approximately linearly with increasing
temperature reaching a blocking temperature of $\rm T_b$ =
200$^{\circ}$C in this film. The slope of the decrease in H$\rm _E$
with temperature is approximately the same for all of the films
measured, indicating that the room temperature exchange field is a
good predictor of $\rm T_b$. H$\rm _c$ also decreases with increasing
temperature but at a slower rate reaching the room temperature value
of a free NiFe layer (1-2 Oe) at 230$^{\circ}$C.  Temperature cycles
above $\rm T_N$ with subsequent cooling in modest fields reduced H$\rm
_c$ of the NiFe from 84 Oe to 60 Oe, but did not change H$\rm _E$
significantly.

In some bilayer films no offset was produced in the NiFe magnetization
loop by the NiO buffer, but a large room temperature coercivity as
well as a clear uniaxial anisotropy was observed (Hc=35 Oe Hs= 80 Oe
as compared to H$\rm _c$ = 1-2 Oe and H$\rm _s$ = 5 Oe in a free NiFe
layer). This enhanced H$\rm _c$ indicates that interface exchange
coupling is present, but that it averages to
zero.\cite{malozemoff,soeya} The temperature dependence of H$\rm
_{ce}$ in such a film composed of NiO(355\AA)/NiFe(100\AA) is also
shown in Fig. 3. The coercivity drops with increasing temperature
similar to the behavior seen in films with H$\rm _E$ $>$ 0, reaching
1-2 Oe at 130$^{\circ}$C. The hard axis saturation field has similar
temperature dependence. The reduced temperature where H$\rm _{ce}$
goes to zero may be a result of a reduced $\rm T_N$ for the thinner
NiO layer, however it seems more likely that the reduction is linked
in the same way as the lack of H$\rm _E$ to the interfacial
properties.

\section{DISCUSSION}

	The simplest model desribing NiO/NiFe exchange coupling
indicates the highest H$\rm _E$ should be observed when the NiO
surface is oriented to maximize the number of uncompensated spins,
{\em i.e.} the (111) planes.\cite{meiklejohn} We have shown that this
model clearly does not describe NiO/NiFe exchange couples.  The data
represented in Fig. 2 indicate that the bulk texture of the NiO layer
has very little influence on the interface exchange coupling energy
since the exchange field does not correlate with the texture of the
NiO films. Further, Fig. 1b shows a strong H$\rm _E$ for NiFe films
grown on the (001) thus should produce zero exchange field.  The
substrate properties influence H$\rm _E$, however, since H$\rm _E$ for
the bilayer on MgO is roughly one third that of the polycrystalline
film deposited simultaneously on an amorphous buffer layer.

	Our results agree with Lai {\em et al.}\cite{lai} who use an
MOCVD technique to grow epitaxial (001) oriented NiO buffer layers.
They also find non-zero H$\rm _E$ values for NiFe grown on (001)
oriented NiO, which are roughly half what they measure for NiFe grown
on polycrystalline NiO buffers. However, they measure isotropic
coercivities which are nearly an order of magnitude larger than those
of reactively sputtered or IBS NiO films. This difference in
coercivity may be due to the elevated surface roughness of the
epitaxial MOCVD NiO layers relative to those deposited using the other
techniques.  Moran and coworkers\cite{moran} describe NiFe layers
deposited on single crystal CoO substrates and shows that rougher
interfaces or interfaces with more crystalline disorder produce higher
exchange fields (CoO has structural and magnetic properties similar to
NiO).  Growth studies of IBS films indicate the broad distribution of
adatom energies present in IBS deposition can produce very smooth
surfaces and interfaces.\cite{martin} Low-angle XRD data on the IBS
grown NiO/NiFe bilayers show the surface has approximately 5\AA\ of
roughness.

	In summary, models for the bilayer magnetic response should
divide the NiO into two layers.  We can estimate the thickness of a
surface layer in the NiO, whose spins are dynamic during the NiFe
reversal and so are responsible for the large NiFe coercivity, is
equal to the critical NiO thickness needed to produce the
unidirectional exchange anisotropy. A static NiO layer below the
dynamic layer establishes the direction of $\rm H_E$.  Large $\rm H_E$
are observed independent of the texture of the NiO layer. Thus the
interfacial interaction of the NiFe and the NiO is not influenced by
the average NiO morphology.  Unfortunately it is difficult to probe
the properties of the interface independently of the rest of the NiO
film and observe the magnetic structures that form.  Variations in
H$\rm _c$ and $\rm H_E$ in coupled bilayers as well as the presence of
a critical thickness, and the training effect, indicate that the
dynamics of the interfacial antiferromagnetic domain structure of the
NiO during the NiFe magnetization reversal is the key to understanding
the magnetic response of oxide based exchange
couples.\cite{malozemoff}

   	The antiferromagnetic order of the NiO and the presence of
non-zero $\rm H_E$ is unaffected by bulk morphological variations. In
this sense, the NiO is a more forgiving exchange bias layer than
FeMn/NiFe or NiMn/NiFe exchange couples, since these compounds require
the specific FCC structure throughout the buffer layer to produce the
antiferromagnetic phase and achieve a non-zero $\rm H_E$.\cite{lin2}
This difference may be due to the more robust antiferromagnetism
associated with the super-exchange interaction and ionic bonding in
the oxide materials.

	We would like to thank Keith Wilfinger for helpful discussions
and R. G. Musket for RBS measurements.  Part of this work was
performed under the auspices of the U. S.  Department of Energy (DOE)
by LLNL under contract No. W-7405-ENG- 48, and part was supported by
DOE's Tailored Microstructures in Hard Magnets Initiative. LEJ
supported by DOE's Science and Engineering Research Semester, and
Partners in Industry and Education programs.  Manuscript submitted
March 1, 1996.

\clearpage


\clearpage

\begin{figure}
\caption{Magnetization as a function of applied field for two
NiO(500\AA)/NiFe(100\AA) films deposited simultaneously on different
substrates are shown.  a) shows the easy axis and hard axis response
for the film deposited on an amorphous $\rm Al_2O_3$ film. b) shows the
response in the in- plane (100) directions on the (001) face of MgO,
parallel and perpendicular to $\rm H_b$.}
\end{figure}

\begin{figure}
\caption{Interface exchange energy, J = $\rm H_E M_s t_{NiFe}$, as a
function of the x-ray intensity ratio of the NiO(111) to the NiO(200)
reflections. The plot shows there is no correlation between the
texture of the NiO buffer layer and the resulting exchange field.}
\end{figure}

\begin{figure}
\caption{Exchange field and coercive field for two NiO/NiFe coupled
bilayer films as a function of temperature.  The circles show HE and
Hc for a strongly exchange biased bilayer. The triangles show Hc for a
bilayer with zero exchange field but elevated coercivity due to the
interface coupling.  HE goes to zero at the blocking temperature, $\rm T_b$,
which is lower than the bulk Neel phase transition temperature, $\rm
T_N$ ($\rm T_N$ = 250$^{\circ}$C for NiO).}
\end{figure}


\begin{references}

\bibitem[\dag]{Remail} michel@cmsgee.llnl.gov

\bibitem[*]{Aemail} chaiken@llnl.gov

\bibitem[\ddag]{Yemail} ykim@tdh.qntm.com

\bibitem[\P]{Lemail} lantz\_johnson@internetqm.llnl.gov


\bibitem{matt} M. J. Carey and A. E. Berkowitz, "Exchange anisotropy
in coupled films of NiFe with NiO and CoNiO," Appl. Phys. Lett.,
vol. 60, pp. 3060-3062, 1992.

\bibitem{matt2} M. J. Carey, F. E. Spada, A. E. Berkowitz, W. Cao and
G. Thomas, "Preparation and structural characterization of sputtered
CoO, NiO, and NiCoO thin epitaxial films," J. Mater. Res., vol. 6,
pp. 2680-2687, 1991.

\bibitem{kennedy} R. J. Kennedy, "The growth of iron oxide, Nickel
oxide and cobalt oxide thin films by laser ablation from metal
targets," IEEE trans. Magn., vol. 31, pp. 3829-3831, 1995.

\bibitem{lind} D. M. Lind, S. D. Berry, G. Chern, H. Mathias and
L. R. Testardi, "Growth and structural characterization of $\rm Fe_3O_4$ and
NiO thin films and superlattices grown by oxygen-plasma-assisted
molecular-beam epitaxy," Phys. Rev. B, vol. 45, pp. 1838-1850, 1992.

\bibitem{lai} C.-H. Lai, H. Matsuyama, R. L. White and T. C. Anthony,
"Anisotropic exchange for NiFe films grown on epitaxial NiO," IEEE
Trans. Magn., vol.  31, pp. 2609-2611, 1995.

\bibitem{fesiprb} A. Chaiken, R. P. Michel and M. A. Wall, "Structure
and magnetism of Fe/Si multilayers grown by ion-beam sputtering,"
Phys. Rev., vol. B53, pp. 5518-5541, 1996.

\bibitem{lin} C.-L. Lin, J. M. Sivertsen and J. H. Judy, "Magnetic
properties of NiFe films exchange-coupled with NiO," IEEE Trans. Mag.,
vol. 31, pp. 4091- 4093, 1995.

\bibitem{malozemoff} A. P. Malozemoff, "Mechanisms of exchange
anisotropy," J. Appl.  Phys., vol. 63, pp. 3874-3879, 1988.

\bibitem{soeya} S. Soeya, T. Imagawa, K. Mitsuoka and S. Narishige,
"Distribution of blocking temperatures in bilayered NiFe/NiO films,"
J. Appl. Phys., vol. 76, pp. 5356-5360, 1994.

\bibitem{meiklejohn} W. H. Meiklejohn and C. P. Bean, "New Magnetic
Anisotropy," Phys.  Rev., vol. 105, pp. 904-913, 1956.

\bibitem{moran} T. J. Moran, J. M. Gallego and I. K. Schuller,
"Increased exchange anisotropy due to disorder at permalloy/CoO
interfaces," J. Appl. Phys., vol.  78, pp. 1887-1891, 1995.

\bibitem{martin} P. J. Martin and R. P. Netterfield, "Optical films
produced by ion-based techniques," in Progress in Optics, vol. 23,
E. Wolf, Eds. North Holland, 1986, pp. 114-182.

\bibitem{lin2} T. Lin, C. Tsang, R. E. Fontana and J. K. Howard,
"Exchange-coupled Ni-Fe/Fe-Mn, Ni-Fe/Ni-Mn, and NiO/NiFe films for
stabilization of magnetoresistive sensors," IEEE Trans. Mag., vol. 31,
pp. 2585-2590, 1995.

\end{references}
\end{document}